\documentclass[a4paper,aip,jcp,reprint,superscriptaddress,amsmath,amssymb]{revtex4-1}

\usepackage{graphicx,color}

\begin{document}

\title{The effect of boundary adaptivity on hexagonal ordering and bistability in circularly confined quasi hard discs}

\author{Ian Williams}
\email{ian.williams@bristol.ac.uk}
\affiliation{H.H. Wills Physics Laboratory, Tyndall Ave., Bristol, BS8 1TL, UK}
\affiliation{School of Chemistry, Cantock's Close, University of Bristol, BS8 1TS, UK}
\affiliation{Centre for Nanoscience and Quantum Information,
Tyndall Avenue, Bristol BS8 1FD, UK}
\author{Erdal C. O\u{g}uz}
\affiliation{Institut f\"ur Theoretische Physik II, Heinrich-Heine-Universit\"at, D-40225 D\"usseldorf, Germany}
\author{Robert L. Jack}
\affiliation{Department of Physics, University of Bath, Bath, BA2 7AY, UK}
\author{Paul Bartlett}
\affiliation{School of Chemistry, Cantock's Close, University of Bristol, BS8 1TS, UK}
\author{Hartmut L\"owen}
\affiliation{Institut f\"ur Theoretische Physik II, Heinrich-Heine-Universit\"at, D-40225 D\"usseldorf, Germany}
\author{C. Patrick Royall}
\affiliation{H.H. Wills Physics Laboratory, Tyndall Ave., Bristol, BS8 1TL, UK}
\affiliation{School of Chemistry, Cantock's Close, University of Bristol, BS8 1TS, UK}
\affiliation{Centre for Nanoscience and Quantum Information,
Tyndall Avenue, Bristol BS8 1FD, UK}

\date{\today}

\begin{abstract}
\textbf{The behaviour of materials under spatial confinement is sensitively dependent on the nature of the confining boundaries. In two dimensions, confinement within a hard circular boundary inhibits the hexagonal ordering observed in bulk systems at high density. Using colloidal experiments and Monte Carlo simulations, we investigate two model systems of quasi hard discs under circularly symmetric confinement. The first system employs an adaptive circular boundary, defined experimentally using holographic optical tweezers. We show that deformation of this boundary allows, and indeed is required for, hexagonal ordering in the confined system. The second system employs a circularly symmetric optical potential to confine particles without a physical boundary. We show that, in the absence of a curved wall, near perfect hexagonal ordering is possible. We propose that the degree to which hexagonal ordering is suppressed by a curved boundary is determined by the `strictness' of that wall.}
\end{abstract}

\maketitle

\section{Introduction}

The behaviour of materials under spatial confinement is modified compared to that in the bulk. On reducing a system to a lengthscale comparable to the size of its constituent particles one observes new structures \cite{oguz2009,barreirafontecha2008,ha2004}, modified dynamics \cite{eral2009,edmond2012} and phase behaviour differing from that of the bulk system \cite{pieranski1983,schmidt1996,schmidt1997}. 

While many effects of confinement are due to finite system size and volume exclusion effects near a wall, others can be attributed to the properties of the boundary. For instance, local density is enhanced in the vicinity of an attractive wall and decreased near a repulsive wall \cite{devirgiliis2007}. As a consequence, the freezing temperature of materials is raised with respect to the bulk when confined by attractive walls while repulsive walls have the opposite effect, lowering the freezing temperature \cite{christenson2001,albasimionesco2006}. Furthermore, confinement by smooth boundaries affects behaviour differently than similar rough walls. For sufficient system density, smooth walls induce particle layering resulting in oscillatory density profiles perpendicular to the boundary, often promoting crystallisation \cite{donnelly2002,ohnesorge1994,stratton2009,haghgooie2006,mittal2008}. If instead the wall is rough on the particle lengthscale, this layering is inhibited and crystallisation is suppressed \cite{teboul2002,sarangapani2008,eral2009}. Dynamically, rough walls are found to suppress particle mobility compared to smooth walls resulting in an increase in relaxation time \cite{nemeth1999,scheidler2002,scheidler2003}. The density (in colloidal samples) or temperature (in molecular liquids) at which the glass transition occurs is similarly dependent on the confining lengthscale and the boundary details \cite{christenson2001,nugent2007,edmond2010,nemeth1999,barut1998}. Through understanding and controlling the boundaries confining a system one can alter the energy landscape it experiences, offering new routes to self-assembly and control of reaction rates and crystal growth \cite{ha2004,hamilton2011,curk2012,zhou2008}.

The research described here is concerned with a two-dimensional hard disc system. At sufficient density, bulk hard discs adopt a locally hexagonal structure \cite{bernard2011,alder1962}. When confined to a narrow channel by smooth, hard walls, however, new structural ordering emerges. For confining walls separated by a distance corresponding to an integer number of close packed particle layers, the confinement is commensurate with the crystalline lattice and hexagonal ordering is unimpeded \cite{rice2009,stratton2009}. In the incommensurate case, when an integer number of hexagonal layers do not fit the channel, buckled structures arise \cite{lowen2010}. At lower particle densities, a modulated fluid structure exists, consisting of particle layers parallel to the walls \cite{chaudhuri2008}.

Although locally hexagonal ordering is possible in hard disc systems confined by smooth, flat walls, such ordering is incommensurate with a curved boundary, resulting in the suppression of hexagonal ordering in a variety of systems confined by a hard circular wall. Instead, circularly confined samples adopt a concentrically layered structure that mimics the symmetry of the confining geometry \cite{nemeth1998,bubeck1999,bubeck2002,watanabe2011} as predicted by density functional theory \cite{kim2001}. If, instead of a strictly hard wall, the circular boundary is soft or adaptive then qualitatively different behaviour is observed. We have previously shown that a deformable circular boundary capable of responding to the shape of a confined sample allows locally hexagonal ordering for sufficiently dense systems \cite{williams2013}. The result is an entropically-driven bistability between concentrically layered configurations reminiscent of hard wall confinement and these structures with enhanced locally hexagonal order reminiscent of bulk hard discs at comparable densities. Such locally hexagonal ordering is not observed in equivalent systems confined by hard walls, suggesting that this behaviour is facilitated by the decreased `strictness' of a deformable boundary compared to a hard wall.

Here we explore the role of the strictness of a curved boundary in inhibiting locally hexagonal ordering in circularly confined hard disc systems through a combined experimental and simulated approach to two model systems. We describe the relationship between boundary deformation and hexagonal ordering in the experimental model system known as the `colloidal corral' \cite{williams2013} and show, via Monte Carlo simulation, that controlling the deformability of the confining boundary allows the tuning of the aforementioned structural bistability. A qualitative comparison is made to a second model system for soft circularly symmetric confinement defined without a wall, representing a minimally strict boundary. A similar comparison has previously been employed by Schweigert \textit{et al.} \cite{schweigert2000} to study re-entrant behaviour in the melting of two-dimensional clusters. We demonstrate that decreasing boundary strictness is inherently associated with the capacity for enhanced locally hexagonal ordering as long as the boundary is sufficient to maintain the system at high density. The implication is that the structure of a confined material can be altered by modifying the boundary, allowing \emph{in situ} driving of a system between qualitatively distinct configurations.

\section{Model Systems}
\label{secModelsystems}

\subsection{The Colloidal Corral}

\begin{figure}[htb]
\begin{center}
\centerline{\includegraphics[natwidth=80mm,natheight=32mm]{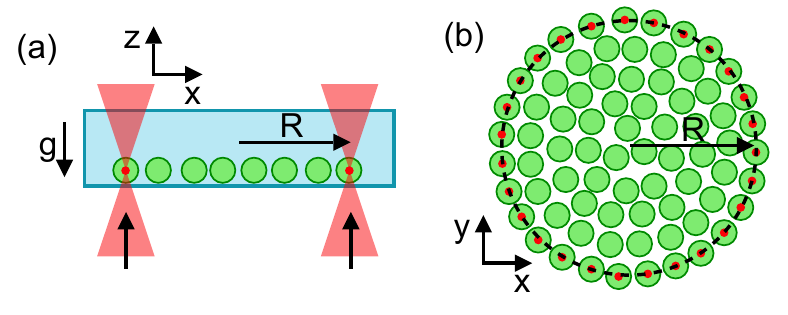}}
\caption{\label{figCorralschematic} Schematic showing colloidal corral geometry from two perspectives. (a) Cross-sectional view in the $x$--$z$ plane highlighting gravitational confinement to quasi-two-dimensions. (b) Top-down view in the $x$--$y$ plane, as seen in micrographs. Particles marked with red dots are localised in optical traps. $R$ is the radius of the optically defined confining ring. }
\end{center}
\end{figure}

\textit{Experimental Details --- } The colloidal corral model system is defined in a quasi-two-dimensional colloidal suspension by localising $27$ particles on a circle using holographic optical tweezers \cite{liesener2000,bowman2013}. These optically trapped particles create a deformable circular boundary for additional identical particles confined to the interior. A schematic of this geometry is shown in Fig. \ref{figCorralschematic}. The experimental colloidal sample consists of polystyrene spheres of diameter $\sigma = 5 \; \mathrm{\mu m}$ and polydispersity $s = 2 \%$ in a water--ethanol mixture at a ratio of $3:1$ by weight. The density mismatch between the particles and the solvent is such that their gravitational length is $l_g / \sigma_{\mathrm{eff}} = 0.015(1)$ resulting in sedimentation of suspended particles and the formation of a quasi-two-dimensional monolayer adjacent to a glass coverslip. This coverslip is made hydrophobic by treatment with Gelest Glassclad 18 to prevent particle adhesion.

An optically trapped sphere displaced from its potential minimum experiences a Hookean restoring force and as such the corral boundary is deformable and capable of responding to the interior sample in an adaptive manner. Corral adaptivity is characterised by a single radial spring constant, $\kappa_{\mathrm{exp}}$, determined by measuring the probability distribution of radial co-ordinates for boundary particles in the absence of a confined population. The optical potential is extracted and fit with the parabolic form characteristic of a Hookean spring. The resulting spring constant is $\kappa_{\mathrm{exp}} = 302(2)$ $k_BT \sigma_{\mathrm{eff}}^{-2}$ where $\sigma_{\mathrm{eff}}$ is the effective hard sphere diameter of the polystyrene particles, defined below. Experimental data are acquired for up to 6 hours at $0.5$ frames per second and particle trajectories are extracted \cite{crocker1996}.

\textit{Simulation Details --- } Complementary Monte Carlo simulations of a similarly confined hard disc system are performed. $27$ discs are located in parabolic potential energy wells arranged on a circle of radius $R_0$, mimicking the optical traps employed experimentally. The ratio of the corral radius and the disc diameter serves as a fit parameter for matching simulation to experiment. Best agreement is found for $R_0/\sigma = 4.32$. This fit is then used to determine the effective Barker-Henderson hard sphere diameter for the experimental system \cite{barker1976} which accounts for electrostatic interactions between colloids. This results in a Debye length of $\lambda_{\mathrm{D}}^{-1} \approx 25 \mathrm{nm}$ which is consistent with the experimental conditions. In this manner, the effective hard sphere diameter is determined to be $\sigma_{\mathrm{eff}}=5.08 \; \mathrm{\mu m}$. Both experimental and simulated data are reported in terms of the internal effective area fraction defined as $\phi_{\mathrm{eff}}= (\pi \sigma_{\mathrm{eff}}^2)/(4 \langle A_{\mathrm{Vor}} \rangle)$ where $\langle A_{\mathrm{Vor}} \rangle$ is the average area per particle calculated via a Voronoi decomposition of the system.

\begin{figure}[htb]
\begin{center}
\centerline{\includegraphics[natwidth=70mm,natheight=70mm]{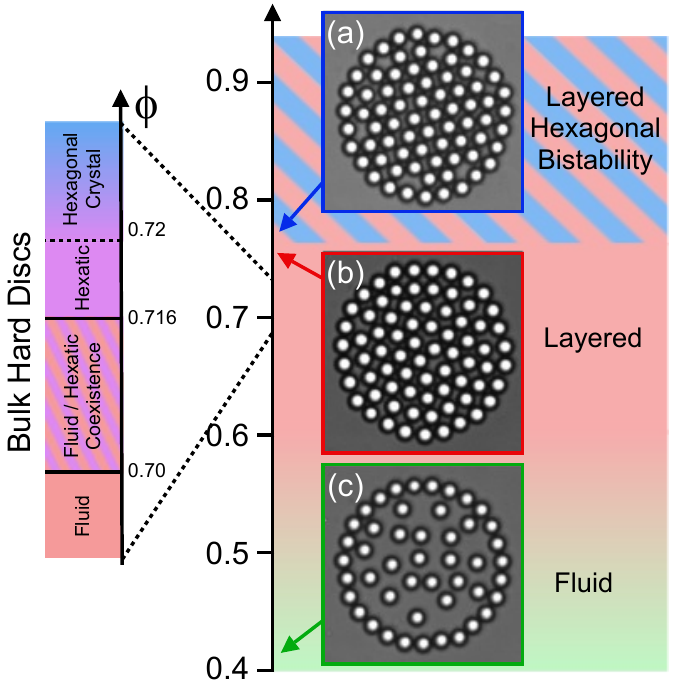}}
\caption{\label{figPhasediagram} Colloidal corral phase diagram and illustrative micrographs. At low density the interior is fluid-like (c). On increasing density a concentrically layered structure develops (b). For $\phi_{\mathrm{eff}} \gtrsim 0.77$ there exists a bistability between concentrically layered and locally hexagonal configurations (a). Left hand side shows phase behaviour for bulk hard discs at comparable area fractions \cite{bernard2011}.}
\end{center}
\end{figure}

\textit{Corral Phase Behaviour --- } We have previously reported the phase behaviour observed in both the experimental and simulated corral system \cite{williams2013} which is summarised in Fig. \ref{figPhasediagram}. At low effective area fraction the interior sample is fluid-like as shown in Fig. \ref{figPhasediagram} (c). On increasing the confined population the curved wall forces the formation of concentric particle layers [Fig. \ref{figPhasediagram} (b)]. This structure is qualitatively similar to those observed for circular confinement imposed by a hard wall \cite{nemeth1998,bubeck1999,bubeck2002,watanabe2011}. For $\phi_{\mathrm{eff}} \gtrsim 0.77$, however, a bistability is observed between concentrically layered structures and configurations exibiting a greater degree of locally hexagonal ordering [Fig. \ref{figPhasediagram} (a)]. 

\subsection{The Optical Bowl}
\textit{Experimental Details --- } As a point of comparison with both the adaptive corral confinement described above and the hard wall circular confinement described in the literature \cite{nemeth1998,bubeck1999,bubeck2002,watanabe2011,kim2001,schweigert2000} a second experimental model system is presented enabling circularly symmetric confinement in the absence of a physical boundary. The holographic optical tweezers system used in forming the colloidal corral employs a $100 \times$ magnification microscope objective of numerical aperature $NA = 1.3$. This focuses the trapping laser tightly, resulting in stable optical tweezers. By replacing this lens with an objective of numerical aperature $NA = 0.5$ the beam is only weakly focused. Such an arrangement is identical to that employed by Arthur Ashkin in his initial observations of laser radiation pressure \cite{ashkin1970} and is insufficient to stably trap a colloidal sphere in three dimensions. Laterally, a dielectric particle exposed to such a weakly focused beam experiences an optical gradient force acting towards the beam axis while it is axially accelerated downstream by the optical scattering force. 

\begin{figure}[htb]
\begin{center}
\centerline{\includegraphics[natwidth=80mm,natheight=75mm]{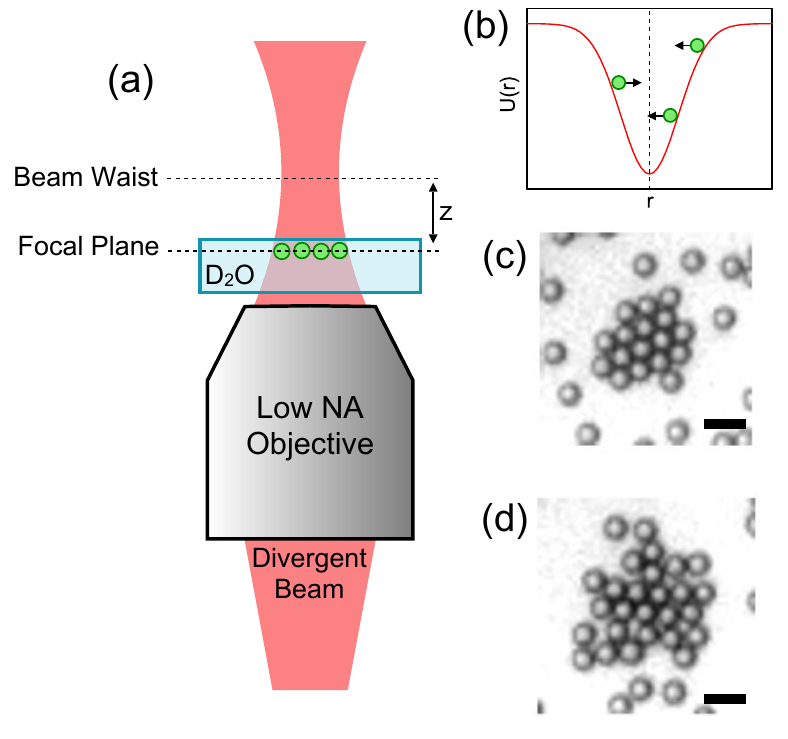}}
\caption{\label{figBowlschematic} (a) Schematic illustrating optical system forming an extended, bowl-like optical potential capable of exerting gradient forces on multiple particles simultaneously. (b) 1 dimensional schematic of the optical bowl potential well capable of acting on multiple particle simultaneously. Arrows indicate forces acting towards the minimum on the optical axis. (c, d) Example micrographs showing clusters of varying size and quality assembled within optical potentials created in this manner. Scale bars indicate $10 \; \mathrm{\mu m}$.}
\end{center}
\end{figure}

By replacing the water--ethanol solvent mixture with deuterated water, the density mismatch forming quasi-two-dimensionality in the colloidal corral experiments is inverted and the polystyrene particles cream to the top of the sample cell. The optical scattering force acting on these creamed particles has no effect as it is entirely balanced by the microscope slide substrate resulting in a quasi-two-dimensional colloidal sample that is free to explore the optical energy landscape defined by the lateral gradient forces. For the loosely focused Gaussian beam employed here this landscape consists of a central energy minimum on the beam axis and is everywhere else attractive towards this minimum. The loose focus also results in a beam waist that is larger than the particle diameter, meaning the optical potential energy landscape is capable of exerting forces on multiple particles simultaneously. All particles illuminated by the beam are attracted towards the beam axis, resulting in an extended, bowl-like optical potential. This setup is illustrated schematically in Fig. \ref{figBowlschematic}.

The width and depth of this potential are controlled by altering the beam divergence at the objective entrance aperture and the laser power respectively. A divergent beam is focused downstream of the objective focal plane, subjecting the sample to a potential of greater width than the beam waist as illustrated in Fig. \ref{figBowlschematic} (a). Increasing divergence increases the potential width. Similarly, an increase in laser power results in a deeper potential.

The optical bowl is circularly symmetric and acts upon multiple particles simultaneously, drawing them towards the beam axis. A one dimensional illustration of this is shown in Fig. \ref{figBowlschematic} (b). Consequently, two-dimensional clusters are observed in the potential region, two examples of which are depicted in Fig. \ref{figBowlschematic} (c) and (d). The size and degree of hexagonal ordering within a cluster are dependent on the width and depth of the applied potential. Experiments are performed in which clusters assemble until they span the potential region, and the final assembled states are analysed. The assembly of similar two-dimensional colloidal clusters has previously been reported by Ju\'{a}rez \textit{et al.} employing an alternating quadrupolar electric field in creating a bowl-like energy landscape \cite{juarez2012lc,juarez2012afm}. 

\textit{Simulation Details --- } Monte Carlo simulations are performed in which a dilute hard disc system is allowed to explore a Gaussian potential well defined by its depth and standard deviation. In contrast with the parabolic well investigated by Schweigert \textit{et al.} \cite{schweigert2000}, the optical bowl is assumed to be Gaussian due to the Gaussian profile of the laser beam employed in experiment. As with the experimental system, the simulated clusters are allowed to evolve until they span the potential and their final structures are analysed. Five independent simulations are performed at every combination of width and depth considered. By matching the properties of the experimentally observed clusters to those formed in simulation the experimental optical potentials are characterised in terms of their depth and width. 

\subsection{Structural Analysis}

In both model systems hexagonal ordering about a given particle $j$ is quantified using the bond orientational order parameter, $\psi_6^j$, defined as
\begin{equation}
\label{eq:localpsi6}
\psi_{6}^{j}  = \left| \frac{1}{z_j} \sum_{m=1}^{z_j} \exp{(i 6 \theta_m^j)} \right|
\end{equation}
where $z_j$ is the co-ordination number of particle $j$, $m$ labels its neighbours and $\theta_m^j$ is the angle made between a reference axis and the bond joining particles $j$ and $m$. The vertical bars represent the magnitude of the complex exponential. Particle neighbours are identified through a Voronoi decomposition. $\psi_6^j = 1$ represents perfect hexagonal ordering around particle $j$ while lower values represent a weaker degree of local hexagonal structure. Taking the average of $\psi_6^j$ over all confined particles yields an average bond-orientational order parameter, $\psi_6$, characterising the hexagonality of the entire system. 

In the corral system, boundary curvature inhibits hexagonal ordering in the particle layer directly adjacent to the wall. In all experimental samples for which $\phi_{\mathrm{eff}} > 0.7$, $\psi_6^j \approx 0.5$ in this layer. Since this layer contains up to half of the confined population this wall-defined value of $\psi_6^j$ dominates the spatial averaging, suppressing variations in $\psi_6^j$ deeper in the corral. For this reason, $\psi_6^j$ averages within the colloidal corral are taken over particles whose radial position $r < 0.7 R$, where $R$ is the corral radius. This excludes the wall-adjacent layer from the average, rendering $\psi_6$ more sensitive to changes in local structure.

Clusters formed in the optical bowl are identified using a connectivity algorithm based on a Voronoi decomposition with the condition that neighbouring particles must be separated by $< 1.3 \; \sigma_{\mathrm{eff}}$. This method is insensitive to reasonable changes in this neighbour cut-off length \cite{malins2013}. Hexagonal ordering within the cluster is characterised by averaging $\psi_6^j$ over all cluster particles.

\section{Results}
\subsection{Deformation of the Colloidal Corral}

\begin{figure*}[htb]
\begin{center}
\centerline{\includegraphics[natwidth=160mm,natheight=59mm]{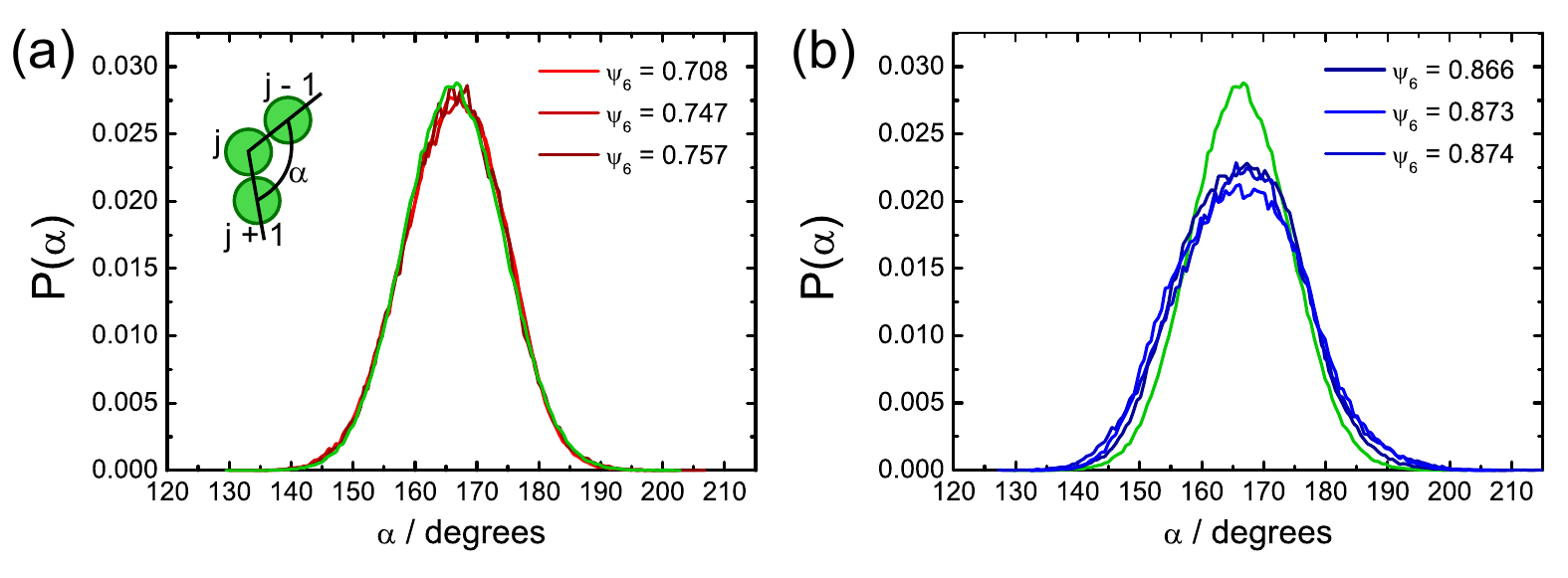}}
\caption{\label{figWallangles} Experimental boundary angle distributions for samples in the bistable region ($\phi_{\mathrm{eff}} > 0.77$, population $N \ge 47$) corresponding to (a) low $\psi_6$, concentrically layered structures and (b) high $\psi_6$ locally hexagonal configurations. Each distribution corresponds to a single experimental sample labelled by the time-averaged $\psi_6$. Inset in (a) shows a section of the corral wall illustrating that lines joining the centres of adjacent wall particles make an angle $\alpha$ at particle $j$. In both panels the green line corresponds to the boundary angle distribution for the unpopulated system.}
\end{center}
\end{figure*}

We have previously suggested that the observation of locally hexagonal structures within the colloidal corral is due to boundary adaptivity \cite{williams2013}. Here we explicitly show that these configurations cause significant deformation of the corral wall compared to concentrically layered structures at similar effective area fraction and population. 

The $27$ optical traps defining the corral boundary are equispaced on a circle, and thus sit at the vertices of a regular $27$-sided polygon, the interior angles of which are $\alpha = 180 - (360 / 27) = 166.\dot{6}^{\circ}$. The instantaneous corral shape is the $27$-sided polygon defined by the locations of the $27$ optically trapped particles. For an unpopulated corral in the absence of Brownian motion, this polygon coincides with the regular $27$-sided polygon and thus its interior angles are all $\alpha = 180 - (360 / 27) = 166.\dot{6}^{\circ}$. However, since optically trapped particles are free to explore their local environment due to Brownian motion, we observe a distribution of internal angles. By comparing the distribution of $\alpha$ for a populated corral to that of an unpopulated corral, boundary distortion is characterised.

Figure \ref{figWallangles} shows typical experimentally measured boundary angle distributions, $P(\alpha)$. In both (a) and (b) the green line represents this distribution for the unpopulated corral and is peaked at $\alpha \approx 166^{\circ}$ as anticipated for a regular $27$-sided polygon. Red lines in Fig. \ref{figWallangles} (a) show the internal angle distributions for low $\psi_6$, concentrically layered samples in the bistable region ($\phi_{\mathrm{eff}} > 0.77$). The distributions show little deviation from the unpopulated distribution, suggesting that concentrically layered structures cause minimal deformation in the corral wall. This is contrasted with Fig. \ref{figWallangles} (b) which shows similar data for high $\psi_6$, locally hexagonal samples. Here the boundary angle distribution is qualitatively modified such that it no longer coincides with the unpopulated distribution, showing that locally hexagonal structures result in significant deformation of the corral boundary. The mean of the distribution is unchanged as the internal angles of the $27$-sided polygon always sum to $4500^{\circ}$, but its width and symmetry about this mean are modified demonstrating that, on average, the corral wall must adopt an altered shape in order to accommodate a configuration with strong locally hexagonal ordering.

Distortions to the corral boundary shape are further characterised by considering the second and third moments of the boundary angle distribution. The second moment, or variance, characterises the width of a distribution and is defined as:
\begin{equation}
\label{eqVariance}
V = \sum (\alpha - \bar{\alpha})^2 P(\alpha)
\end{equation}
where $\bar{\alpha}$ is the mean of $P(\alpha)$ and $V$ has dimensions of degrees squared. The third moment, or skew, characterises the symmetry of a distribution about its mean and is defined as:
\begin{equation}
\label{eqSkew}
\gamma = \frac{1}{V^{3/2}} \sum (\alpha - \bar{\alpha})^3 P(\alpha)
\end{equation}
and is dimensionless.

\begin{figure*}[htb]
\begin{center}
\centerline{\includegraphics[natwidth=160mm,natheight=60mm]{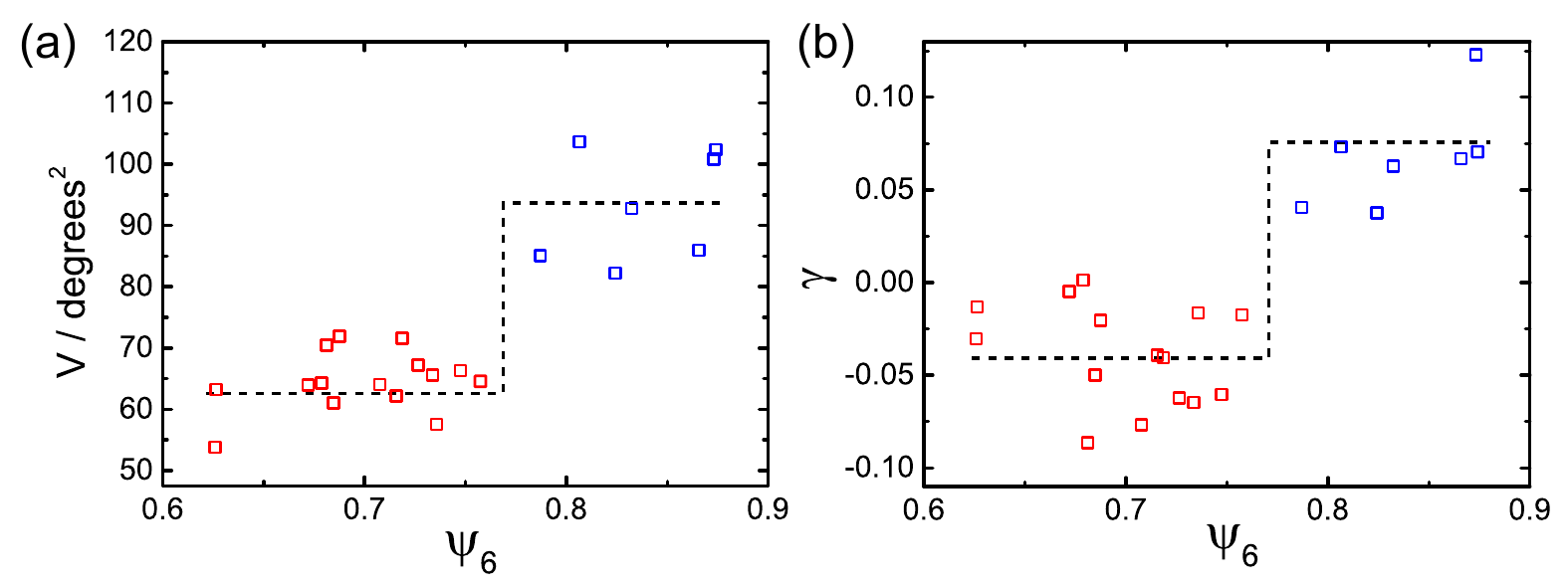}}
\caption{\label{figVarianceskew} (a) Variance and (b) skew of experimentally measured corral boundary angle distributions as a function of time-averaged $\psi_6$ for confined populations $44 \le N \le 49$. Red points represent concentrically layered samples and blue points show locally hexagonal configurations. Lines guide the eye, indicating sharp transitions in variance and skew at $\psi_6 \approx 0.775$.}
\end{center}
\end{figure*}

Figure \ref{figVarianceskew} shows (a) the variance and (b) the skew of the experimentally measured corral boundary angle distributions as a function of time-averaged $\psi_6$. Data are coloured based on the structures identified experimentally with red points representing concentric layering and blue points representing locally hexagonal configurations. In both cases, two clouds of data are evident separated by an abrupt transition at $\psi_6 \approx 0.775$. Locally hexagonal structures result in a wider and positively skewed distribution of boundary angles, reinforcing that significant corral deformation is required for their observation. Here we reiterate that equivalent locally hexagonal structures are not observed in systems confined by a hard circular boundary, and thus we conclude that adaptivity or deformability is a requirement for this local hexagonal ordering that is reminiscent of bulk hard discs.

Despite locally hexagonal configurations deforming the adaptive wall, the overall elastic energy in the boundary is in fact reduced when compared to concentrically layered structures of the same internal population. While hexagonal structures require a small number of local boundary distortions, the overall packing density within the corral is increased compared to layered structures meaning that, on average, the corral can contract and wall particles tend to sit closer to their optical energy minima. For layered structures, the entire corral is isotropically stretched, preserving the shape of the wall angle distribution but increasing the corral radius. This is supported by our mechanical pressure measurements within the corral \cite{williams2013}.

\subsection{Controlling Structure in Circular Confinement}

\begin{figure*}[htb]
\begin{center}
\centerline{\includegraphics[natwidth=160mm,natheight=62mm]{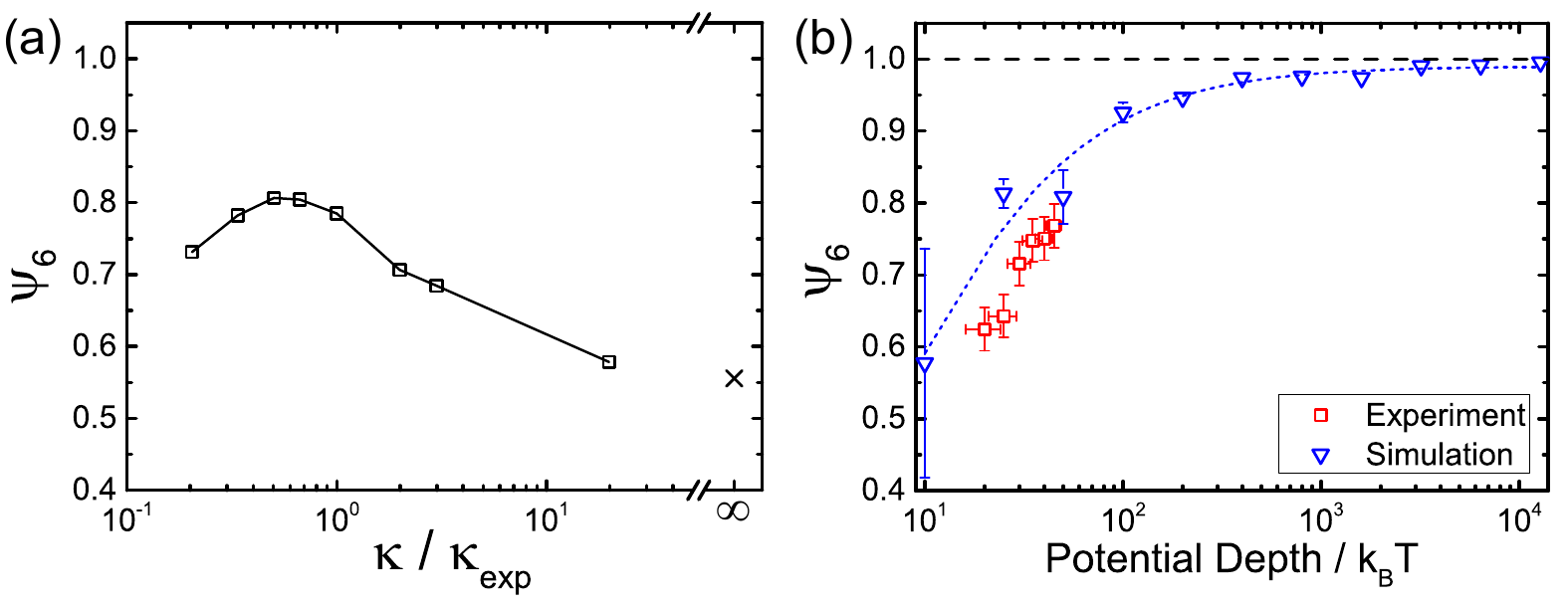}}
\caption{\label{figPsi6compare} (a) Average $\psi_6$ in simulated corral systems of population $N=49$ as a function of boundary spring constant, $\kappa$ (open squares). Spring constants are given in terms of the experimental spring constant $\kappa_{\mathrm{exp}} = 302(2)$ $k_BT \sigma_{\mathrm{eff}}^{-2}$. Cross at $\kappa = \infty$ indicates $\psi_6$ of the best known packing for $N=49$ discs in hard circular confinement \cite{graham1998}. (b) Average $\psi_6$ for clusters formed in the optical bowl. Red points represent experimental data, blue points are from Monte Carlo simulation. Data is averaged over $5$ independent simulations or $3$ independent experiments. Blue dashed line serves to guide the eye. Uncertainty in $\psi_6$ is the standard error in the mean. Uncertainty in experimental potential depths are due to matching the experimental potentials to simulated potentials.}
\end{center}
\end{figure*}

\textit{The Colloidal Corral --- } Local hexagonal ordering in the colloidal corral requires boundary deformation. A softer wall incurs a smaller energetic cost for a given deformation suggesting that hexagonal ordering may be more prevalent if the boundary spring constant is reduced. This is considered to be a decrease in the strictness of the confining geometry. In Monte Carlo simulation, the corral may be defined using arbitrarily high or low spring constants, facilitating the study of systems that are inaccessible to experiment.

Figure \ref{figPsi6compare} (a) shows the effect of altering this stiffness on the average $\psi_6$ measured in the simulated confined system of population $N=49$. Increasing the spring constant from $\kappa_{\mathrm{exp}}$ results in the suppression of hexagonal ordering characterised by a decrease in $\psi_6$. Thus, we show that a stiffer or less adaptive boundary inhibits local hexagonality in systems confined by a curved wall. As $\kappa \to \infty$ the adaptive boundary behaves similarly to a hard wall, and as such the data for high spring constant is considered as approximating the layering behaviour expected. The best known packings of monodisperse discs within a hard circular boundary for populations in the range $44 \le N \le 49$, as calculated by Graham \emph{et al.} \cite{graham1998}, have area fractions in the range $0.78 < \phi < 0.791$, corresponding to bistable area fractions in the colloidal corral. Hexagonal ordering in these best known packings is restricted to $0.4 < \psi_6 < 0.6$, which is comparable to the degree of hexagonal ordering observed in corral simulations with spring constant $\kappa = 20 \kappa_{\mathrm{exp}}$.

Decreasing the spring constant from the experimental conditions results initially in a small increase in $\psi_6$ coming to a maximum at $\kappa \approx 0.5 \kappa_{\mathrm{exp}}$. Further decrease in wall stiffness below $\kappa \approx 0.5 \kappa_{\mathrm{exp}}$ causes $\psi_6$ to again decrease. This apparent inhibition of hexagonal ordering for very low corral spring constants is due to expansion of the system. In such weak parabolic potentials, wall particles can be located a long way from their energy minima incurring only a small energetic penalty compared to the thermal energy, thus the boundary is unable to maintain the corral interior at high density and the system expands, resulting in low $\psi_6$.

This pattern of low $\psi_6$ for very weakly and very strongly defined walls separated by an intermediate region of enhanced $\psi_6$ for intermediate boundary adaptivities is reminiscent of the non-monotonic yield often observed in self assembling systems \cite{grant2011,klotsa2011,whitelam2009}. The quality of assembly is strongly protocol dependent \cite{klotsa2013,royall2011,royall2012}. Weak interactions (\emph{i.e.} very low spring constants) are insufficient to allow self-assembly and strong interactions (\emph{i.e.} very high spring constants) frustrate ordering, both of which lead to poor assembly. Good assembly is only obtained in an intermediate region of interaction strength, analogous to the observation of highly hexagonal structures only for intermediate spring constants. Spring constants in this region result in a corral wall that is sufficiently stiff that confinement is well-defined but sufficiently adaptive that the boundary deformation required for locally hexagonal ordering is possible.

\textit{The Optical Bowl --- } Comparison is now drawn to the clusters formed in the optical bowl, confined without a curved boundary. Figure \ref{figPsi6compare} (b) shows the average $\psi_6$ within fully assembled clusters in both experiment (red points) and simulation (blue points) as a function of potential depth. Clusters are considered to be fully assembled, when they cease increasing in size. Experimentally, this occurs within 1 hour of assembly. Data are averaged over $5$ independent simulations or $3$ independent experiments. Considering initially only the experimental data, it is evident that increasing potential depth results in increased hexagonal ordering. A deeper potential exerts greater gradient forces on the particles resulting in stronger confinement. However, experimentally, $\psi_6$ is only observed in the range $0.6 < \psi_6 < 0.8$ --- similar to that observed in the colloidal corral. 

Monte Carlo simulation allows investigation of experimentally inaccessible potential depths, facilitating arbitrarily strong confinement in the absence of a physical boundary. Increasing the potential depth in simulation up to $> 10^{4} k_B T$ shows that, for sufficiently deep bowl-like potentials, near optimal hexagonal ordering is obtained as $\psi_6 \to 1$. This represents an enhancement in locally hexagonal ordering compared to even the optimally hexagonal corral system. There is no evidence of frustration inhibiting cluster assembly for such deep potential wells. The implication of this monotonic increase in $\psi_6$ is that, in the absence of a curved boundary, near perfect hexagonal ordering is possible in a system of hard discs confined to a circular region. 

\textit{Non-monotonicity --- } That $\psi_6$ behaves non-monotonically on increasing the corral spring constant [Fig. \ref{figPsi6compare} (a)] but monotonically on increasing the depth of the optical bowl [Fig. \ref{figPsi6compare} (b)] is explained by qualitative differences between the model systems, specifically the absence of a curved wall in the bowl-like potential well. Increasing the depth of the optical bowl increases the packing density of particles within the assembled clusters, leading to enhanced hexagonal ordering. The initial increase in $\psi_6$ observed in the corral is due to a similar effect --- a circular wall defined with a low spring constant is unable to maintain the confined population at high density, and thus $\psi_6$ is low. Increasing the spring constant contracts the boundary leading to stronger confinement, increased density and thus enhanced hexagonal ordering. This dependence of $\psi_6$ on system density is directly related to the behaviour of hard discs in the bulk \cite{bernard2011} and is the only effect contributing to structural ordering in the optical bowl. The non-monotonicity observed for the corral system is due to the interplay of this density-driven ordering with a second effect --- the inhibition of hexagonal ordering by a curved boundary. Increasing the corral spring constant above $\kappa = 0.5 \kappa_{\mathrm{exp}}$ results in a decrease in $\psi_6$ due to the increasing energetic cost of the boundary deformations required for hexagonal ordering.

For high spring constant the corral boundary behaves similarly to a hard, circular wall and as such the observed structural ordering is compared to the mathematical problem of packing monodisperse discs into a circular region \cite{graham1998,nemeth1998,huang2011,lubachevsky1997}. For the best known packings of identical discs in hard, circular confinement average $\psi_6$ is restricted to the range $0.4 < \psi_6 < 0.6$ due to boundary curvature, which is consistent with $\psi_6$ measured for the highest spring constants considered. In contrast, the optical bowl is more akin to the generation of clusters of monodisperse discs by minimising their second moment about their centroid \cite{graham1990,chow1995}. Such clusters always show strong locally hexagonal ordering, much like the clusters formed in sufficiently deep bowl-like potentials. There is no explicitly curved boundary enclosing an energetically flat interior acting to suppress hexagonal ordering. Similarly, colloidal crystals assembled in approximately parabolic potentials in the absence of a physical curved boundary, as reported by Ju\'{a}rez \textit{et al.}, show a monotonic increase in hexagonality on increasing the applied voltage \cite{juarez2012lc,juarez2012afm}.

\subsection{Tuning Structural Bistability}

For corral confinement under experimental conditions, structural bistability is observed for $\phi_{\mathrm{eff}} \gtrsim 0.77$, corresponding to interior populations $N \geq 47$ \cite{williams2013}. This bistability is characterised by two peaks in the probability distribution of $\psi_6$ corresponding to concentrically layered and locally hexagonal structures. Since Fig. \ref{figPsi6compare} (a) shows that the degree of hexagonal ordering is dependent on boundary adaptivity, it is expected that altering the spring constant will lead to a change in this $\psi_6$ probability distribution.

\begin{figure}[htb]
\begin{center}
\centerline{\includegraphics[natwidth=80mm,natheight=50mm]{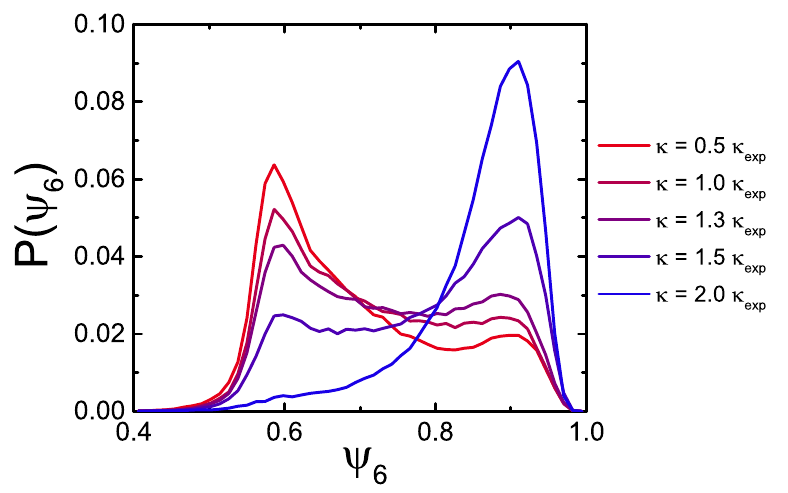}}
\caption{\label{figPsi6pdfkappa} $\psi_6$ probability distributions within the colloidal corral at population $N=47$ obtained via Monte Carlo simulation for varying boundary spring constant. Spring constants are given in terms of the experimental spring constant $\kappa_{\mathrm{exp}} = 302(2)$ $k_BT \sigma_{\mathrm{eff}}^{-2}$.}
\end{center}
\end{figure}

Figure \ref{figPsi6pdfkappa} shows $\psi_6$ probability distributions obtained via Monte Carlo simulation for a range of corral spring constants. The most adaptive boundary considered is defined with $\kappa = 0.5 \kappa_{\mathrm{exp}}$, corresponding to the maximum in Fig. \ref{figPsi6compare} (a). Under these conditions a single peak is observed at $\psi_6 \approx 0.9$ representing strongly locally hexagonal configurations. As the spring constant is increased the boundary is less adaptive and a second peak develops in the vicinity of $\psi_6 \approx 0.6$, corresponding to concentrically layered structures. This two-peak distribution represents the structural bistability. As $\kappa$ is further increased the relative heights of these two peaks changes. The high $\psi_6$ peak becomes weaker and the low $\psi_6$ peak stronger, representing an increased likelihood of observing the confined system in a concentrically layered configuration. This behaviour is also reflected in the average $\psi_6$ plotted in Fig. \ref{figPsi6compare} (a).

These data suggest that the structural bistability observed in both simulation and experiment can be tuned via the spring constant. Thus, the structural ordering of the confined sample can be controlled by altering the properties of the boundary. Increasing the spring constant of a corral boundary (\emph{e.g} by increasing the laser power forming the optical traps) confining a locally hexagonal configuration should drive the interior population into a concentrically layered structure. Similarly, decreasing the spring constant confining a concentrically layered structure should allow local hexagonal ordering in the corral interior as the energetic cost of boundary deformation is decreased. Such driving of the bistability in an experimental system has not yet been attempted. Control over structural ordering in a confined system by altering the adaptivity of the boundary represents controlled, wall-induced freezing and melting.

\section{Discussion \& Conclusion}

\begin{figure}[htb]
\begin{center}
\centerline{\includegraphics[natwidth=80mm,natheight=90mm]{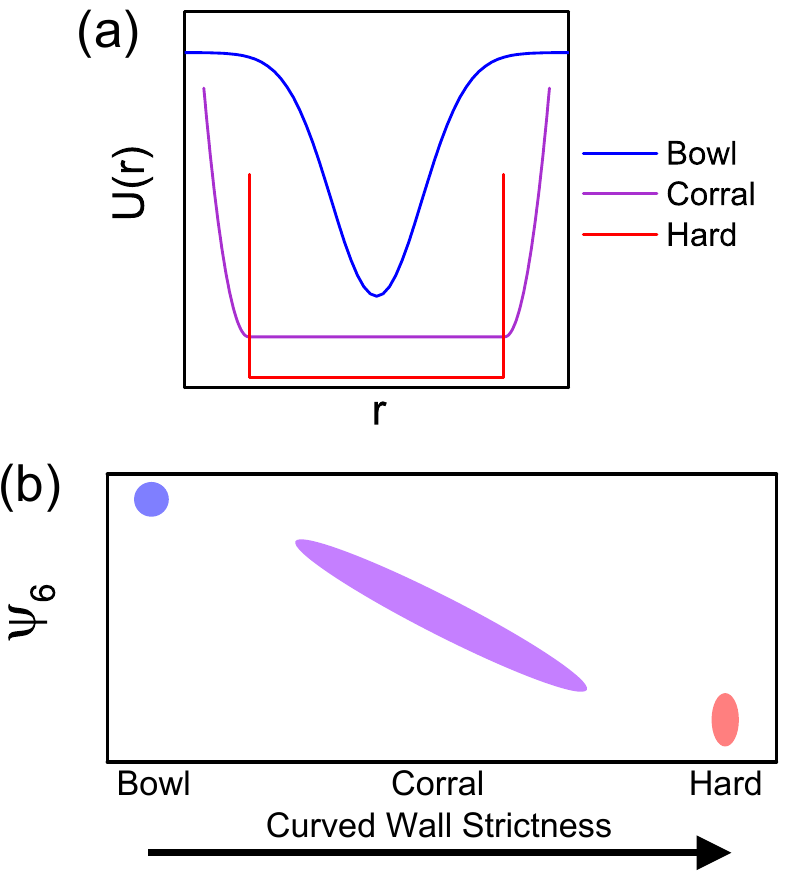}}
\caption{\label{figStrictness} (a) Radial cross-section of the potential energy landscape for optical bowl confinement (blue), adaptive corral confinement (purple) and hard circular confinement (red). (b) Schematic representation of the effect of the strictness of a curved wall on hexagonal ordering in circularly symmetric confinement. The optical bowl (blue) represents minimally strict confinement and allows near perfect hexagonal ordering. Hexagonal ordering in adaptive corral confinement (purple) is dependent on the deformability of the wall (assuming system density is maintained), with a stiffer wall representing a stricter boundary. A hard circular wall (red) inhibits hexagonal ordering.}
\end{center}
\end{figure}

In the bulk, hard disc systems tend to adopt hexagonal local structures at high density. Circular confinement by a hard wall represents the maximally strict circular confining geometry and strongly inhibits this hexagonal ordering except in special cases of magic numbers for confined population \cite{lubachevsky1997,graham1998,huang2011}. In general, a hard circular wall imposes a concentrically layered structure upon a confined two-dimensional sample. The colloidal corral model system represents a reduction in wall strictness by allowing deformations of the boundary. We further propose that the optical bowl represents a minimally strict circular confining geometry as the confining potential is axisymmetric but there is no curved wall maintaining the system and inhibiting hexagonal ordering. The three potentials considered are depicted in Fig. \ref{figStrictness} (a). We suggest that these three cases span a range of wall strictness in which hexagonal ordering becomes more prevalent as strictness is decreased. This is represented schematically in Fig. \ref{figStrictness} (b).

In the presence of a physical curved boundary, locally hexagonal ordering is only permitted when deformations are possible, and as such the colloidal corral shows an increase in $\psi_6$ as its spring constant is reduced and deformations have a lower energetic cost. However, this relationship is non-monotonic as a very low spring constant is insufficient to maintain the confined system at the high density required for hexagonal configurations. We have shown explicitly that high $\psi_6$ configurations result in significant distortions to the corral wall. Furthermore, the bistability between locally hexagonal and concentrically layered configurations is sensitively dependent upon the corral spring constant. Altering the spring constant results in a change in the probability distribution of $\psi_6$, suggesting that control over boundary stiffness facilitates the tuning of the observed bistability and the driving of the system between qualitatively distinct structures.

The bowl-like optical potential confines particles in the absence of a curved boundary, and thus hexagonal ordering is not inhibited by curvature. We have shown that $\psi_6$ in clusters assembled in these potentials increases with potential depth up to $\psi_6 \approx 1$, representing perfect hexagonal ordering.

The combined interpretation of these findings is that wall curvature is the dominant influence in the inhibition of hexagonal ordering and that significant locally hexagonal ordering in confinement requires a reduction of strictness in the boundary. The minimally strict boundary allows perfect hexagonal ordering for sufficiently strong confinement. Intermediate between the hard wall and no wall case is adaptive confinement. We have shown that the degree to which an adaptively confined system can adopt locally hexagonal configurations depends upon the degree to which it can deform its confining boundary.

That freezing and melting in confined systems can be induced by altering the boundary properties has implications in creating microscopic reconfigurable devices in which system properties are controlled externally. Additionally, adaptive boundaries have clear relevance in the modelling of biological systems in which a densely crowded environment is typically enclosed by a flexible membrane \cite{zimmerman1993,elcock2010}. That the structural properties of a crowded system are influenced by boundary adaptivity offers insight into the behaviour of particles confined in deformable containers \cite{maibaum2001} such as emulsion droplets \cite{kim2008,shirk2013} or vesicles \cite{dinsmore1998}. This heralds new possibilities for templated assembly \cite{whitesides2002} and the manufacture of mesoscopic clusters with a variety of structural properties.

\section{Acknowledgements}
CPR and IW gratefully acknowledge the Royal Society for funding. IW was supported by the EPSRC. RLJ was supported by EPSRC grant EP/I003797/1. Financial support from the European Research Council (ERC Advanced grant INTERCOCOS, project number 267499) is acknowledged. We thank David Carberry and Richard Bowman for assistance in setting up the holographic optical tweezers hardware and software respectively. We express our gratitude to Bob Evans and Dave Phillips for useful discussions.

\end{document}